\documentclass[12pt]{article}   
\usepackage{fullpage}
\usepackage[super,sort,comma]{natbib}

\usepackage{amssymb}
\usepackage{amsmath}
\usepackage{hyperref}
\usepackage{algorithm}
\usepackage{algpseudocode}
\usepackage{subfig}
\usepackage{xcolor}
\usepackage{authblk}
\usepackage{graphicx}

\DeclareMathOperator*{\argmin}{arg\,min}

\begin{document}
\newcommand{\Rp}{\mathbb{R}_{\geq 0}}
\newcommand{\Rpp}{\mathbb{R}_{+}}
\newcommand{\pluseq}{\mathrel{+}=}
\newcommand{\asteq}{\mathrel{*}=}

\title{Equivariant Multiscale Learned Invertible Reconstruction for Cone Beam CT: From Simulated to Real Data} 

\author[1,2]{Nikita Moriakov} 
\author[2]{Efstratios Gavves}
\author[3]{Jonathan H. Mason}
\author[1]{Carmen Seller-Oria}
\author[1]{Jonas Teuwen} 
\author[1]{Jan-Jakob Sonke}

\affil[1]{Netherlands Cancer Institute, Plesmanlaan 121, Amsterdam 1066 CX, the Netherlands}
\affil[2]{University of Amsterdam, Science Park 900, Amsterdam 1098 XH, the Netherlands}
\affil[3]{Elekta Limited, Cornerstone, Crawley, UK}

\maketitle

\begin{abstract}
Cone Beam CT (CBCT) is an important imaging modality nowadays, however lower image quality of CBCT compared to more conventional Computed Tomography (CT) remains a limiting factor in CBCT applications. Deep learning reconstruction methods are a promising alternative to classical analytical and iterative reconstruction methods, but applying such methods to CBCT is often difficult due to the lack of ground truth data, memory limitations and the need for fast inference at clinically-relevant resolutions. In this work we propose LIRE++, an end-to-end rotationally-equivariant multiscale learned invertible primal-dual scheme for fast and memory-efficient CBCT reconstruction. Memory optimizations and multiscale reconstruction allow for fast training and inference, while rotational equivariance improves parameter efficiency. LIRE++ was trained on simulated projection data from a fast quasi-Monte Carlo CBCT projection simulator that we developed as well. Evaluated on synthetic data, LIRE++ gave an average improvement of 1 dB in Peak Signal-to-Noise Ratio over alternative deep learning baselines. On real clinical data, LIRE++ improved the average Mean Absolute Error between the reconstruction and the corresponding planning CT by 10 Hounsfield Units with respect to current proprietary state-of-the-art hybrid deep-learning/iterative method.
\end{abstract}

\section{Introduction}
\label{s.intro}
Computed Tomography (CT) is one of the most used medical imaging modalities nowadays. Similar to many other modern imaging modalities such as MRI, the measurements acquired by a CT scanner - i.e., X-ray projection images taken from a multitude of angles - are not immediately usable in clinic and instead need to undergo the process of \emph{reconstruction}, wherein they are processed by a reconstruction algorithm and combined into a three-dimensional volume. An important type of CT is Cone Beam Computed Tomography (CBCT), where the X-ray source emits rays in a wide cone-shaped beam and the detector is a large flat panel array. In CBCT, both the X-ray source and the detector typically follow circular trajectories around the isocenter, and the detector is sometimes offset to give a larger field of view \citep{letourneau2005}. CBCT has applications in interventional radiology \citep{floridi2014}, dentistry \citep{dawood2009} and image-guided radiation therapy \citep{jaffray2002}, however, CBCT image quality remains poor compared to classical CT with helical trajectory for a few reasons. CBCT reconstruction is inherently harder since the data completeness condition for exact reconstruction of the whole volume is not satisfied for circular source/detector orbits \citep{maas2010, tuy1983}. In addition to common CT artifacts such as photon starvation, scattering becomes a particularly prominent issue, since a large detector panel captures more scattered photons from a wide cone beam of X-rays. The poor resulting Hounsfield Unit (HU) calibration is a limiting factor for applications in e.g. adaptive radiotherapy, where a daily CBCT scan with sufficient quality to enable online delineation and treatment plan optimization would be highly desirable \citep{sonke2019}.

Deep learning reconstruction methods have drawn a lot interest from the medical imaging community by achieving remarkable results in public reconstruction challenges such as FastMRI \citep{fastmri2020, beauferris2020}. \textit{Learned iterative schemes} in particular are powerful family of deep learning reconstruction methods, which are inspired by classical iterative methods such as Landweber iteration, and embed the forward operator directly in the neural network architecture. Intuitively, this allows to `learn a prior from the data' instead of an explicit regularization. Learned Primal-Dual (LPD) algorithm \citep{Adler2017b} is a prominent example of a learned iterative scheme inspired by the Primal-Dual Hybrid Gradient (PDHG) method \citep{pdhg2011}, which combines both image-space and projection-space operations in an end-to-end trainable network. Image-space computations are performed by \emph{primal blocks} and projection-space computations are performed by \emph{dual blocks}, all primal/dual blocks being small convolutional neural networks. LPD framework has been extended to other modalities as well, such as Digital Breast Tomosynthesis \citep{teuwen2021} and MRI \citep{ramzi2020a}, but there are also recent examples of learned iterative schemes for CT \citep{airnet} or MRI \citep{yiasemis2022} reconstruction which operate in image domain only.

Despite their benefits, learned iterative schemes are hard to scale up to a fully three-dimensional modality such as CBCT due to memory limitations. For example, for clinically relevant radiotherapy applications a voxel pitch of at most $2$ mm (isotropic) is desirable, since $2$ mm grids are common in radiotherapy dose computations. For a typical patient, such voxel pitch would result in roughly $256 \times 256 \times 256$ CBCT volume. Given a $256 \times 256 \times 256$ FP32 tensor, a \emph{single} convolution layer with $64$ features would already require 8 GB memory to perform the backpropagation operation. One of the first memory-efficient alternatives is $\partial $U-Net \citep{hauptmann2020}, which is a simpler scheme that does not operate in the projection space. Memory usage is reduced by relying on a multiscale approach, where reconstructions obtained at different resolutions are merged together by a final U-Net. iLPD, or invertible learned primal-dual method, has been considered \citep{rudzusika2021}, where it was shown that it substantially reduces memory requirements and allows to use longer learned iterative schemes. For a 3D helical CT setting, iLPD has been combined \citep{rudzusika24} with splitting the scanning geometry in chunks of data that can be processed independently, however, such geometry splitting is not possible for CBCT. To address this issue, LIRE \citep{lire2023} method was recently proposed, where a learned invertible primal-dual scheme was augmented with tiling computation mechanism inside the primal/dual blocks during both training and inference, allowing to use higher filter counts as well as more complex U-Net cells inside primal blocks. LIRE inference takes around 30 seconds on NVIDIA A100 GPU with clinically relevant geometry and resolution, from which it is desirable to speed it up further for future clinical application. A logical step would be to try to combine learned invertible primal-dual scheme and multiscale reconstruction\footnote{It might appear counterintuitive, since the input and the output in a reversible neural network have the same dimensionality, but will be explained in Section \ref{ss.lirep}.}, but it has not been done in literature at the moment, even though invertible flows that incorporate multiscale latent codes are well known in generative modeling \citep{glow2018}.

The task of building natural symmetries of learning tasks into neural network architectures has been a fruitful recent research direction in inverse problems and deep learning in general. For instance, when a patient is rotated we expect the new reconstruction to be a rotated version of the original reconstruction. For convolutional neural networks, this problem is addressed with group equivariant convolutions \citep{cohen2016}, which often allow one to achieve state of the art results at reduced parameter counts on image classification tasks. Group equivariant convolutional neural networks have been applied to inverse problems with learned iterative schemes as well \citep{celledoni2021}, but not in the context of CBCT and learned-primal dual family of methods. 

Regardless of the architectural choices, supervised training with a perfect ground truth knowledge is generally a preferred setting for deep learning-driven reconstruction \citep{jssl2024}. While self-supervised learning (SSL) methods such as Noise2Inverse \citep{hendriksen2020} perform well in fanbeam CT reconstruction, the nature of scatter-induced artifacts makes SSL unsuitable in CBCT setting due to strong correlation of scatter signals in adjacent projections. Furthermore, exact ground truth for the actual patient data remains unknown in CBCT, necessitating the use of a CBCT physics simulator for the generation of synthetic projection data from digital phantoms derived from e.g. patient CT scans. Accurate numerical simulation of CBCT projections is challenging, since both primary and scatter signals depend on material and photon energy distributions, demanding a computationally expensive Monte Carlo procedure. Unfortunately, existing Monte Carlo simulators are not sufficiently fast for use with on-the-fly randomly augmented CT data and are also not well integrated with deep learning APIs such as PyTorch, making development of deep learning reconstruction models on realistic synthetic data difficult for the deep learning community.

In this work we address these challenges and present a novel LIRE++ reconstruction method suitable for reconstruction at 2 mm as well as 1 mm voxel pitches. We start with the development of a fast CBCT physics simulator supporting a per-voxel water-bone material mix, which relies on quasi-Monte Carlo method for scatter estimation \citep{lin21}. The simulator is implemented as a PyTorch CUDA extension. The background on tomography and X-ray scatter physics is provided in the \nameref{ss.physics}.  The inverse problem of CBCT reconstruction is formulated in Section \ref{ss.recon}. The forward model is described in \ref{ss.tomo}, \ref{ss.path}. LIRE++ and the baselines are trained and evaluated using synthetic CBCT projection data generated from a mix of thorax, abdomen, pelvic CTs which serve as ground truth, whereas additional proof-of-concept evaluation is done on real pelvic CBCT projection data. The data is described in more details in Section \ref{ss.data}, and the baselines are given in Section \ref{ss.baselines}. To solve the CBCT reconstruction problem we present LIRE++ model in Section \ref{ss.lirep}. LIRE++ is a fast and parameter-efficient rotationally equivariant multiscale invertible learned primal-dual scheme, which extends LIRE+ from our preliminary unrefereed report \citep{lire2024}. Compared to \citep{lire2023}, the multiscale nature of LIRE++ leads to faster inference and the use of equivariant convolutions improves model robustness. Importantly, LIRE++ was designed to handle real CBCT projection data with large amounts of scatter such as pelvic CBCT acquisitions. Unlike \citep{lire2023, lire2024}, LIRE++ explicitly incorporates scatter correction in end-to-end trainable network and is trained on data with realistically simulated polychromatic primary and scatter signals. The main version of LIRE++ is trained and evaluated for volumes at 2 mm voxel pitch, whereas the additional proof-of-concept version is designed for volumes at 1 mm voxel pitch.

We perform extensive evaluation of LIRE++ and the baselines on synthetic data using image quality metrics such as PSNR and Structural Similarity Index Measure (SSIM), as well as HU Mean Absolute Error (MAE) in Section \ref{s.synthdata}. Additionally, in order to demonstrate that our model translates well to real data, we compare LIRE++, analytical reconstruction with scatter pre-correction and a state-of-the-art proprietary hybrid deep learning/iterative algorithm on pelvic CBCT data from our institution in Section \ref{s.realdata}. In the same section, we also show that LIRE++ can be scaled to produce full-resolution volumes by providing reconstructions from the proof-of-concept LIRE++ version on real pelvic data with 1 mm voxel pitch and compare them to the proprietary method.

\section{Methods}
\label{s.matmethods}
For an introduction to elemtary X-ray physics for tomography, we refer the reader to the \nameref{ss.physics}.

\subsection{Inverse problem of CBCT reconstruction}
\label{ss.recon}
A classical inverse problem in CBCT reconstruction is to determine effective total tissue attenuation coefficient for a multi-spectral beam from the CBCT projection data, corrupted by scatter and other forms of noise. The effective total tissue attenuation can be approximated by $\mu_{\text{tot}, 60}$ at a fixed energy level of $60$ keV, since the energy level of $60$ keV roughly corresponds to the peak in photon energy histograms in both CT and CBCT acquisitions. Therefore, an accurate estimate of $\mu_{\text{tot}, 60}$ would approximate a CT-like image.

We will approach this inverse problem by finding a Bayes estimator \citep{Kaipio2005} parametrized by a neural network trained in a supervised setting. The goal for the Bayes estimator $\hat \mu_{\textrm{Bayes}}$ is to minimize the expected cost
\begin{equation}\label{eq.bayesloss}
L(\hat \mu) = \mathbb E_{(\mu, y) \sim \pi} \ L(\mu, \hat \mu(y))
\end{equation}
over all estimators $\hat \mu$, where $\pi$ is the distribution of pairs $(\mu, y)$ of total attenuation volumes $\mu=\mu_{\text{tot}, 60}$ and the corresponding CBCT projection images $y$ for the underlying anatomies. $L$ is a fixed cost function given by a sum of mean absolute error and a Structural Similarity loss in the image domain (see Eq. \eqref{eq.ploss}) and a mean absolute error in projection domain for scatter correction. The optimal estimator in \eqref{eq.bayesloss} will be chosen from a certain class of neural networks, and minimization of the cost in \eqref{eq.bayesloss} with respect to the parameters $\theta$ of the network $\mathrm{NN}_{\theta}$ will be carried out via minibatch stochastic gradient descent during network training. That is, a training set $\mathcal D_{\text{train}}^{\text{CT}} = \{ \mu: \mu = \mu_{\text{tot}, 60} \text{ a CT volume}\}$ is used to solve the following minimization problem
\begin{equation}
\theta := \argmin_{\theta} \frac 1 {|\mathcal D_{\text{train}}^{\text{CT}}| }\sum_{\mu \in \mathcal D_{\text{train}}^{\text{CT}}} L(\mu, \mathrm{NN}_{\theta}(\widetilde{\mathcal P}(\mu))),
\end{equation}
where $y=\widetilde{\mathcal P}(\mu)$ is synthetic projection data generated from a CT scan $\mu=\mu_{\text{tot}, 60}$, corrupted by Poisson noise and scatter (see Sections \ref{ss.tomo} and \ref{ss.path} respectively). In general, knowing attenuation $\mu_{\text{tot}, 60}$ from a CT scan is not sufficient for an accurate determination of material composition, however, given the simplified water-bone model we will be able to produce an adequate approximation for the purpose of generating synthetic CBCT projections.

\subsection{Primary simulation} 
\label{ss.tomo}
In this work we simulate a common clinical acquisition geometry for a  Linac-integrated CBCT scanner from Elekta \citep{letourneau2005} with a medium field-of-view setting, offset detector, a full $2 \pi$ scanning trajectory and from $432$ to $944$ projections to approximate actual variability in projection counts see in real data. The source-isocenter distance is $1000$ mm and the isocenter-detector plane distance is $536$ mm. The detector is offset by $115$ mm to the side in the direction of rotation to give an increased Field of View. Square detector panel with a side of $409.6$ mm and $256 \times 256$ pixel array is used. Photons from the X-ray source pass through the collimator and the bow-tie filter, and the resulting photon distribution is simulated and stored in a \emph{phase file}.

The total photon count emitted from the source is denoted by $I_{\Sigma}$. To speed up the computation, X-ray energy spectrum is discretized in 10 energy bins $\{[10i \text{ keV}, 10(i+1) \text{ keV})\}_{i=2}^{11}$ with centers at energy levels $E=\{ 25 \text{ keV}, 35 \text{ keV}, \dots, 115 \text{ keV}\}$. To account for the nonuniformity of the photon distribution across the detector, for  $e \in E$ the intensity map $I_{0, e}$ is defined by binning photon distribution from the phase file. Therefore $I_{0, e}(\sigma)$ is the unattenuated X-ray photon count for energy bin centered around $e \in E$ arriving at a detector element $\sigma$. To simulate energy-dependent detector readings, detector response function $\mathrm{resp}$ is approximated as a piecewise linear function such that $r_{20} = 5, r_{60} = 20, r_{120}=10$, where the values between the key points $20 \text{ keV}, 60 \text{ keV}, 120 \text{ keV}$ are computed with linear interpolation \citep{Poludniowski_2009}.

For both primary and scatter simulation it is necessary to derive a water-bone decomposition from a single-energy CT scan, we adopt the approach of \citep{wang2018}. For notation convenience, Hounsfield values $\mathrm{HU}$ are first converted to modified CT numbers by setting $\hat \rho = 0.001 \cdot \mathrm{HU} + 1$. Next, dimensionless constants $\tau_1 = 1.2, \tau_2 = 1.6, \kappa_b = 0.409$ are set. This allows to define water $\rho_w$ and bone $\rho_b$ relative densities as continuous functions of $\hat \rho$ as
\begin{equation}
\rho_w = 
\begin{cases} 
  0 & \hat \rho < \tau_0 \\
  \hat \rho & \tau_0 \leq \hat \rho < \tau_1 \\
  \frac {\tau_1(\tau_2 - \hat \rho)}{\tau_2 - \tau_1} & \tau_1 \leq \hat \rho < \tau_2 \\
  0 & \hat \rho \geq \tau_2 
   \end{cases}
\end{equation}
and
\begin{equation}
\rho_b = 
\begin{cases} 
  0 & \hat \rho < \tau_1 \\
  \kappa_b \frac {\tau_2(\hat \rho - \tau_1)}{\tau_2 - \tau_1} & \tau_1 \leq \hat \rho < \tau_2 \\
  \kappa_b \hat \rho & \hat \rho \geq \tau_2. 
   \end{cases}
 \end{equation}
Therefore, for $m \in \{w, b\}$ and energy $e > 0$ we set $\mu_{\text{tot}, e}^m(x) := \rho_m(x) \overline \mu_{\text{tot}, e}^m$, resulting units being $\mathrm{mm}^{-1}$. We will denote the resulting mapping from a CT scan $\mu=\mu_{\text{tot}, 60}$ into a collection water-bone attenuations for every energy level by $\widetilde \mu$.
 
The \textit{cone-beam transform operator}, or simply the \textit{projection operator}, is defined as an integral operator
\begin{equation}
\mathcal P(\mu)(t, u) = \int_{L_{t, u}} \mu(z) \mathrm d z, 
\end{equation}
where $\mu = \mu(\cdot): \Omega_X \to \mathbb R$ and $L_{t, u}$ is a line from the source to the detector element $u$ at time $t$. $\mathcal P$ is a linear operator, and Hermitian\footnote{For suitably defined $L^2$ function spaces.} adjoint $\mathcal P^*$ of $\mathcal P$ is called the \textit{backprojection operator}. Using the projection operator $\mathcal P$ and the water-bone decomposition operation, we approximate noisy polychromatic primary (i.e., non-scattered) component of the corresponding set of CBCT projections as a finite sum 
\begin{align}
  \label{eq.primarymodel}
  \widetilde{\mathcal P}_p(\mu) &:= \sum\limits_{e \in E} \mathrm{resp}(e) \cdot \text{\texttt{Poisson}}(I_{0, e}  e^{-\mathcal P (\widetilde \mu_{\text{tot}, e})}) = \nonumber \\ 
  &=\sum\limits_{e \in E} \mathrm{resp}(e) \cdot \text{\texttt{Poisson}}(I_{0, e}  e^{-\mathcal P (\widetilde \mu_{\text{tot}, e}^w + \widetilde \mu_{\text{tot}, e}^b)}).
\end{align}
Therefore, $\widetilde{\mathcal P}_p(\mu)$ denotes simulated primary CBCT projection data given the CT scan $\mu = \mu_{\text{tot}, 60}$.

\subsection{Path integral formalism and scatter simulation}
\label{ss.path}
Given the elementary introduction to X-ray physics in \nameref{ss.physics}, we now briefly describe our approach to scatter simulation, which is based on \citep{lin21}. To simplify the notation we assume that the X-ray source and the detector are fixed. On a high level, the scatter signal $y_{\infty}(\sigma)$ recorded at a detector pixel $\sigma$ is expressed as an integral over the space of possible photon paths. Similarly, for $n \in \mathbb N$ we let $y_n(\sigma)$ denote the scatter contribution from photons that have undergone exactly $n$ Compton or Rayleigh interactions. Then $y_{\infty} = \sum\limits_{n > 0} y_n$, but in practice it suffices to truncate this sum due to rapidly decreasing contribution from higher-order events.

Let $\mathbb R_+^n$ be the set of all non-increasing $n$-tuples of positive numbers. The finite-order \emph{scatter path space} is defined for $n \in \mathbb N$ as
\begin{equation}
  \Pi^n = \{\overline{ x_0x_1x_2 \dots x_n x_{n+1} } : |x_ix_{i+1}| = l_i > 0 \text{ for all }i\} \times \mathbb R_+^{n+1}
\end{equation}
and the infinite-order scatter path space is defined as $\Pi^{\infty} = \cup_{n>0} \Pi^n$. The element $(\mathbf x, \mathbf e) \in \Pi^n$ describes a possible scattered photon path where $x_1,\dots,x_n$ are scattering points and $e_0, \dots, e_{n}$ are photon energies. Scattered paths form a subset of all possible photon paths, which carries a probability measure $\Xi$ expressing the probability of each photon path starting from the X-ray source with known initial distribution. Then the expected scatter signal contribution for the detector element $\sigma$ is formally given by
\begin{equation}
\label{eq.pointsample}
y_{\infty}(\sigma) = I_{\Sigma} \int_{\Pi^{\infty}} \mathbb I(\mathbf x \text{ terminates at } \sigma) \mathrm{resp}(\text{final } e\in \mathbf e) \mathrm d \Xi (\mathbf x, \mathbf e). 
\end{equation}
Similar equation holds for $y_n$ with $\Pi^{\infty}$ replaced by $\Pi^n$. To compute $y_n$, we express this integral as
\begin{align}
\label{eq.pathint}
  &y_n(\sigma)= \nonumber \\
I_{\Sigma}&\int\limits_{S^2 \times \Rpp} \mathrm d \nu_0 (\vec v_0, e_0) \int\limits_{\Rpp} \mathrm d \lambda (l_1 | x_0, \vec v_0, e_0)\int\limits_{S^2 \times \Rpp} \mathrm d \nu(\vec v_1, e_1|x_1, \vec v_0, e_0)  \nonumber  \\
  &\int\limits_{\Rpp} \mathrm d \lambda (l_2 | x_1, \vec v_1, e_1)\int\limits_{S^2 \times \Rpp} \mathrm d \nu(\vec v_2, e_2|x_2, \vec v_1, e_1) \cdots \int\limits_{\Rpp} \mathrm d \lambda (l_n | x_{n-1}, \vec v_{n-1}, e_{n-1}) \nonumber \\
  &\int\limits_{\mathrm{proj}_{x_n}(\sigma) \times \Rpp}  p_0(x_n, \vec v_n, e_n)  \mathrm{resp}(e_n) \mathrm d \nu(\vec v_n, e_n|x_n, \vec v_{n-1}, e_{n-1}).
\end{align}
In the integral above, $\nu_0$ is a probability measure on $S^2 \times \Rpp$ determining the source photon distribution. Next, for a photon $\gamma$ with starting position $x \in \mathbb R^3$, a direction vector $\vec v \in S^2$ and energy $e \in \Rpp$ the measure $\lambda (\cdot | x, \vec v, e )$ specifies photon travel distance distribution defined in \eqref{eq.lambdameas}. To specify direction and energy of a scattered photon, we use the measure $\nu(\cdot |x, \vec v, e)$ on $S^2 \times \Rpp$derived from differential cross-section data. Finally, $\mathrm{proj}_{x}(\sigma) \subset S^2$ denotes the central projection of a detector element $\sigma$ onto the unit sphere with center $x$, i.e., $\mathrm{proj}_{x}(\sigma) := \{ \vec v \in S^2: (x + \Rpp \cdot \vec v) \cap \sigma \neq \varnothing\}$, and $p_0(x_n, \vec v_n, e_n)$ denotes the probability that $\gamma$ escapes the patient defined in \eqref{eq.beer}.

For computational reasons, the integral in \eqref{eq.pathint} cannot be evaluated with quadrature rules alone, making quasi-Monte Carlo methods relevant. In our approach we first rely on quasi-Monte Carlo to sample variables in this integral up to $l_n$ to generate paths $(\mathbf x^1, \mathbf e^1), \dots, (\mathbf x^N, \mathbf e^N) \in \Pi^{n-1}$ for the Compton/Rayleigh events up to order $n$ which are stored in GPU memory. Then, for each $i=1,\dots,n$ and $j=1,\dots, N$ expected scatter contribution from the $i$-th interaction point $x_i^j \in \mathbf x^j$ of $j$-th path for each detector element $\sigma$ is explicitly aggregated. Compared to \citep{lin21}, we explicitly support multiple materials in a single voxel and split scatter simulation into sampling step and integration step. Since both steps are highly parallelizable, this allows for an efficient CUDA implementation.

The path sampling step is presented in Alg. \ref{alg:scatter}, and the integration step is presented in Alg. \ref{alg:scatter2} in the \nameref{ss.physics}. We start with a small set of source photons $\mathit{Src}$, for which we compute expected total scatter signal. An important distinction from classical Monte Carlo is the use of Sobol sequences instead of i.i.d. uniform pseudo-random numbers to sample photon paths in Lines 7, 15, 16, 18, 19 of Alg. \ref{alg:scatter}. In particular, in Lines 15-16 material and interaction are sampled types from the corresponding Bernoulli distributions, in Lines 7 and 19 interaction distances are sampled and in Line 18 scattered photon direction is sampled (which determines scattered photon energy as well). Since we only use $|\mathit{Src}|$ photons from the entire phase file, the output of $\mathrm{Integrate}$ should be scaled by $\frac {I_{\Sigma}}{|\mathit{Src}|}$ to produce total scatter estimate with correct intensity. Therefore, complete scatter simulation procedure for an $i$-th projection can be written as
\begin{equation}
  \label{eq.scattermodel}
  \widetilde{\mathcal P}_s(\mu)(i) := \frac {I_{\Sigma}}{|\mathit{Src}|}  \texttt{Integrate}(\texttt{SamplePath}(\mathrm{Rot}_{\varphi_i}(\mathit{Src}), \widetilde \mu), \widetilde \mu),
\end{equation}
where $\mathrm{Rot}_{\varphi_i}$ denotes the transformation which rotates source photons' initial coordinates and directions to match the X-ray source position with angle $\varphi_i$.

\subsection{Data preparation}
\label{ss.data}

To train and evaluate our model on synthetic CBCT data, we used a combined dataset of 424 thorax CT scans and 50 pelvic CT scans with isotropic axial spacing of $0.7-1.17$ mm and z-axis spacing of either $1$ or $2$ mm. Both datasets had axial slice of $512 \times 512$ voxels. All data was binned to give approximately isotropic $2$ mm voxel pitch, resulting in volumes with fixed size of $256^3$ voxels after padding or cropping. No denoising was applied to the CT scans, since unsupervised denoising could blur very fine details such as fissures leading to over-optimistic image quality metrics. The thorax CT dataset was split into a training set of 260 scans, a validation set of 22 scans and a test set of 142 scans. The pelvic CT dataset was split into training set of 39 scans, validation set of 1 scan and testing set of 9 scans. During training, pelvic data was oversampled to balance the frequency of pelvic and thorax data. For an additional proof-of-concept evaluation on real data, planning CT and CBCT acquisions with corresponding baseline reconstructions were collected for 5 pelvic patients. Pelvic CBCT data was acquired on a Linac-integrated CBCT scanner from Elekta \citep{letourneau2005} with a medium field-of-view setting. Study approval was granted by the IRB of our institute, IRBd20-008.

During model training the projection count is randomly uniformly chosen from $432$ to $944$, whereas during evaluation it is set to $720$, and the photon count per $\text{mm}^2$ is randomly chosen from $16000$ to $66000$ in order to represent a variety of photon counts seen in thorax and pelvic CBCT acquisitions, whereas during evaluation we set thorax photon count to $16000$ and pelvic photon count to $66000$. The intensity maps $I_{0,e}$ and the total photon count $I_{\Sigma}$ in Sections \ref{ss.tomo}, \ref{ss.path} are scaled accordingly. Given the low spatial frequency of the scatter signal, we found it sufficient to simulate scatter at one quarter of the primary pixel pitch, additionally, the scatter is simulated for each eighth projection only. Linear interpolation is used is to upscale simulated scatter to full primary resolution and projection count.

In order to retrieve attenuation information from the raw data recorded by a scanner it is necessary to perform some form of projection normalization, which in practice is accomplished by using gain files which correspond to `air-only' acquisitions. Therefore, given a CT scan $\mu = \mu_{\text{tot}, 60}$, we simulate normalized negative log-transformed projection data
\begin{equation}
\label{eq.normraw}
y_{\text{raw}}(\mu) = - \log \min \left( \frac{\widetilde{\mathcal P}_s(\mu) + \widetilde{\mathcal P}_p(\mu)}{\widetilde{\mathcal P}_p(\text{air})}, 1 \right)
\end{equation}
and normalized negative log-transformed primary projection data
\begin{equation}
\label{eq.normprimary}
y_{\text{primary}}(\mu) = - \log \min \left( \frac{\widetilde{\mathcal P}_p(\mu)}{\widetilde{\mathcal P}_p(\text{air})}, 1 \right).
\end{equation}
The functions $\widetilde{\mathcal P}_p, \widetilde{\mathcal P}_s$ above are defined in \eqref{eq.primarymodel} and \eqref{eq.scattermodel} respectively.

\subsection{Baseline methods}
\label{ss.baselines}
We rely on the following classical baselines for evaluation on synthetic data: FDK \cite{feldkamp84}, PDHG \citep{pdhg2011} with Total Variation (TV) regularisation. As deep learning baselines, we used U-Net \citep{cicek2016} with  FDK reconstruction as input and  $\partial$U-Net \citep{hauptmann2020} with scatter-corrected projection data and scatter-corrected FBP reconstruction as input. Additionally, for thorax data we finetune pre-trained versions of LIRE and LIRE+ from \citep{lire2023, lire2024}, which were developed on the same set of thorax CTs. In all these baselines, we used a two-dimensional U-Net similar to the one in LIRE++ from Section \ref{ss.lirep} for scatter pre-correcton without gradient propagation from reconstruction to the scatter pre-correction step. Our implementation of $\partial$U-Net relies on the open-source implementation\footnote{Adapted to 3D and our projector/backprojector code from \url{https://github.com/asHauptmann/multiscale}} from the author, where the base filter count was increased from 12 to 32 in order to get closer to the base filter counts used by LIRE++ to make the comparison fair while fitting into memory budget. As input to $\partial$U-Net, we provided the scatter pre-corrected FDK reconstruction and the field-of-view tensor $V$ defined later in Section \ref{ss.lirep}. The same augmentation strategy as LIRE++ and the same loss function (see Section \ref{ss.lirep}) were used. To train U-net and $\partial$U-Net, Adam optimizer \citep{kingma2014} was employed with batch size of $8$ on NVIDIA Quadro RTX 8000 cards via gradient accumulation, initial learning rate of $0.0001$ and a plateau scheduler with linear warm-up and 10 epoch patience. The best-performing model on the validation set was chosen for testing.

For the proof-of-concept evaluation on real data, we used FDK with deep-learning scatter pre-correction  and a proprietary commercial hybrid deep learning/iterative method currently employed in our center, which we will refer to as TV++. TV++ utilizes a U-net for scatter pre-correction in the projection domain  and a variation of the Polyquant method from \cite{Mason_2017}. Additional proprietary corrections for glare and detector lag are applied in TV++ as well.

\subsection{LIRE++}
\label{ss.lirep}

\label{ss.arch}
\begin{algorithm}
\centering
\begin{algorithmic}[1]
\Procedure{\texttt{reconstruct}}{$y_{\text{raw}}, \mathcal P, \mathcal P^*, \theta, V, w$}
\State $y \gets \text{\texttt{GC-UNet}}_{\theta^s}(y_{\text{raw}})$ \Comment{Initial scatter correction}
\State $x \gets \text{\texttt{FDK}}_w(y)$ \Comment{FDK initialization for $x$}
\State $\overline x \gets \text{\texttt{Downsample}}_{25\%}(x)$ \Comment{Downsample $x$}
\State $x_{\text{bp}} \gets  \text{\texttt{Downsample}}_{25\%}(\mathcal P^*(w y))$ \Comment{Backproj. scatter-corr. data}
\State $\overline y \gets \text{\texttt{ProjDown}}_{25\%}(y)$ \Comment{Downsample \& subsample projections}
\State $I \gets []$ \Comment{Initialize output list}
\State $f \gets \overline [\overline x, x_{\text{bp}}, \overline x, x_{\text{bp}}, \overline x, x_{\text{bp}}, \overline x, x_{\text{bp}}] \in X^{8}$\Comment{Initialize primal vector}
\State $h \gets \overline y^{\otimes 8} \in Y^{8}$\Comment{Initialize dual vector}
\For{$(i, \alpha) \gets (1, 25\%), (2, 50\%), (3, 100\%)$}
\State $\overline x \gets \text{\texttt{Downsample}}_{\alpha}(x)$ \Comment{Downsample $x$ to current resolution}
\State $\overline y, \overline y_{\text{raw}} \gets \text{\texttt{ProjDown}}_{\alpha}(y), \text{\texttt{ProjDown}}_{\alpha}(y_{\text{raw}})$ \Comment{Down. proj.}
\State $\overline V \gets \text{\texttt{Downsample}}_{\alpha}(V)$ \Comment{Downsample FoV tensor}
\State $\overline w \gets \text{\texttt{Downsample}}_{\alpha}(w)$ \Comment{Downsample weighting tensor}
\State $d_1, d_2 \gets \text{\texttt{Splt}}(h)$ \Comment{Split dual channels}
\State $p_1, p_2 \gets \text{\texttt{Splt}}(f)$ \Comment{Split prime channels}
\State $p_{\text{op}} \gets \mathcal P_{\alpha}([p_2, \overline x]^{\oplus})$ \Comment{Project $p_2$ and $\overline x$}
\State $d_2 \gets d_2 + \Gamma_{\theta_i^d}([p_{\text{op}}, d_1, \overline y,  \overline y_{\text{raw}}]^{\oplus})$ \Comment{Upd. $d_2$}
\State $b_{\text{op}} \gets \mathcal P_{\alpha}^*(\overline w d_2)$ \Comment{Weighted backproj. $d_2$}
\State $\text{\textit{LW}} \gets \mathcal P_{\alpha}^* (\mathcal P_{\alpha} (\overline x) - \overline y)$ \Comment{Landweber term}
\State $p_2 \gets p_2 + \Lambda_{\theta_i^p}([b_{\text{op}}, p_1, \overline x, \text{\textit{LW}}, \overline V]^{\oplus})$ \Comment{Upd. $p_2$}
\State $h \gets [d_1, d_2]^{\oplus}$ \Comment{Combine new dual}
\State $f \gets [p_1, p_2]^{\oplus}$ \Comment{Combine new primal}
\State $x \gets x + \text{\texttt{Upsample}}_{\alpha^{-1}}(\text{\texttt{Conv3d}}(f, \theta_i^o))$ \Comment{Update reconstruction}
\State $I \gets I + [x]$ \Comment{Append new $x$ to output list}
\State $h \gets \text{\texttt{Perm}}(h, {\theta_i^m})$ \Comment{Permute dual channels w. $\theta_i^m$}
\State $f \gets \text{\texttt{Perm}}(f, {\theta_i^m})$ \Comment{Permute prim. channels w. $\theta_i^m$}
\If{$i<3$}
\State $f, h \gets \text{\texttt{Upsample}}_{200\%}(f), \text{\texttt{Upsample}}_{200\%}(h)$ \Comment{Upsample latents}
\EndIf
\EndFor
\State \textbf{return} $y, I$
\EndProcedure
\end{algorithmic}
\caption{LIRE++}
\label{alg:liremain}
\end{algorithm}

LIRE++ method is an unrolled learned iterative scheme, which extends LIRE by relying on a multiscale reconstruction strategy to improve the inference speed, equivariant primal cells for higher parameter efficiency and robustness to orientation, as well as forced centered weight normalization \citep{karras2024} to improve convergence stability. Similar to LIRE, the memory footprint of LIRE++ is reduced by combining invertibility for the network as a whole and patch-wise computations for local operations. An optional CPU-GPU memory streaming mechanism is implemented, which would keep entire primal/dual vectors in CPU memory and only send the patch required for computing the primal/dual updates or gradients into the GPU. We refer the reader to the original work \citep{lire2023} for the discussion on invertibility and patch-wise computations. To justify the combination of multiscale reconstruction and invertibility, we make the following observation: if $\Lambda: \mathbb R^n \to \mathbb R^n$ is an invertible neural network and $\iota: \mathbb R^n \to \mathbb R^m, m \geq n$ is some fixed injective differentiable mapping such as nearest upsampling operation, then the input $x \in \mathbb R^n$ can be restored from the output $\iota(\Lambda(x)) \in \mathbb R^m$ unambiguously by first inverting $\iota$ and then $\Lambda$, so the gradients for the parameters of $\Lambda$ can be computed without storing the activations during the forward pass. The algorithm was implemented as a C++/CUDA extension for PyTorch \citep{pytorch} in order to maximize memory efficiency, training and inference speed.

LIRE++, given by function \texttt{RECONSTRUCT}($y_{\text{raw}}, \mathcal P, \mathcal P^*, \theta, V, w$) in Algorithm~\ref{alg:liremain}, consists of 3 iterations and uses primal/dual latent vectors with 8 channels. Here $y_{\text{raw}}$ is normalized log-transformed and scaled raw projection data from \eqref{eq.normraw}, $\mathcal P$ and $\mathcal P^*$ are normalized projection and backprojection operators respectively, $\theta$ is a list of parameters, $w$ is a projection-domain redundancy weighting for offset detector and $V$ is an auxiliary Field-of-View tensor defined as
\begin{equation*}
V(p) := \frac {\text{number of projections where voxel } p \text{ is visible}} {\text{total projection count}} 
\end{equation*}
The parameters $\theta$ are partitioned into 5 parameter groups, where $\theta^s$ are parameters of the $\text{\texttt{GC-UNet}}$ U-Net for scatter pre-correction, $\{ \theta_i^p \}_{i=1}^{3}$ are the primal block parameters, $\{ \theta_i^d \}_{i=1}^{3}$ are the dual block parameters, $\{ \theta_i^o \}_{i=1}^{3}$ are the output convolution parameters and $\{ \theta_i^m \}_{i=1}^{3}$ are the permutation parameters. For every $i$, the permutation $\theta_i^m$ is some fixed permutation of $[1,2,\dots,8]$ which is randomly initialized during model initialization and stored as a model parameter; we require that $\theta_i^m$ mixes the first and the second half of $[1,2,\dots,8]$. Channel-wise concatenation of tensors $z_1, z_2, \dots, z_k$ is denoted by $[z_1, z_2, \dots, z_k]^{\oplus}$, conversely, function $\text{\texttt{Splt}}(z)$ splits tensor $z$ with $2n$ channels into two halves along the channel dimension. Function $\text{\texttt{Perm}}(z, \rho)$ permutes tensor $z$ with $n$ channels along the channel dimension with the permutation $\rho \in \text{\texttt{Sym}}(n)$. Function $\text{\texttt{Upsample}}_{\alpha}(z)$ performs nearest upsampling of $z$ to $\alpha$ percentage of the resolution, $\text{\texttt{Downsample}}_{\alpha}(z)$ downsamples tensor $z$ to $\alpha$ percentage of the resolution via average pooling and function $\text{\texttt{ProjDown}}_{\alpha}(z)$ downsamples projection tensor $z$ to $\alpha$ percentage of the resolution and drops all but every $1 / \alpha$-th projection. For resolution $\alpha \in [25\%, 50\%, 100\%]$, we write $\mathcal P_{\alpha}, \mathcal P_{\alpha}^*$ for the projection and backprojection operator respectively at $\alpha$ resolution, where for $\alpha$-percentage of resolution only every $1 / \alpha$-th projection is computed.
 
LIRE++ starts with $\text{\texttt{GC-UNet}}$, a residual gradient-checkpointed U-Net for scatter pre-correction. $\text{\texttt{GC-UNet}}$ consists of $5$ layers of two-dimensional convolutions with initial filter count of $16$. Gradient checkpointing mechanism erases all internal activations during the forward pass, which are recomputed during the backprogation. Scatter-corrected projections $y$ are used to produce the initial reconstruction $x$ by applying FDK with redundancy weighting $\text{\texttt{FDK}}_w$. We stress that the initial $x$, as well as all intermediate reconstructions produced by LIRE++, are at full resolution. In contrast with \citep{hauptmann2020, lire2024}, LIRE++ keeps the reconstruction at full resolution and operates by adding corrections at $25\%, 50\%,100\%$ resolution respectively. This multi-scale correction strategy is based on the intuition that the initial high-resolution FDK volume can be efficiently corrected by removing large-scale artifacts at low resolution first and then refining the result at medium and full resolutions, minimizing the total block count and the associated compute costs. Importantly, this strategy is compatible with reversible primal/dual updates.

LIRE++ is built from a number of convolutional blocks. $\text{\texttt{Conv3d}}(\cdot, \theta_i^o)$ denotes a $1 \times 1 \times 1$ convolution with parameters $\theta_i^o$. $\Gamma_{\theta_i^d}$ denotes $i$-th dual block with parameters $\theta_i^d$ comprised of 3 layers of $3 \times 3 \times 3$ convolutions with 64, 64 and 4 filters respectively and LeakyReLU activation after the first and the second convolution layers. $\Lambda_{\theta_i^p}$ denotes $i$-th primal block with parameters $\theta_i^p$, which is a U-Net of depth 1 comprised of 6 convolution layers of $3 \times 3 \times 3$ P4-equivariant convolutions with 48 filters in top layers and 96 filters in the bottleneck. LeakyReLU activations are used after all but the final convolutional layer. Input to a primal block has $8$ channels and no `group dimension', whereas output of the last convolution has $4 \times 4$ channels due to the extra `group dimension'. This output is then averaged over the group dimension, making the primal block equivariant w.r.t. the action of P4 (i.e., $90$-degree rotations along the z-axis). Forced weight normalization \citep{karras2024} is used for the primal/dual block parameters to improve training stability.

The algorithm returns complete scatter-corrected projection data $y$ and a list $I = [x_1, x_2, x_3]$ of reconstructions. The image-domain loss function $L_p$ is a weighted sum of a mean absolute error $\| \cdot \|$ and a SSIM loss, which are taken separately over the full field of view region (i.e., voxels present in at least half of the projections) and the partial field of view region (i.e., voxels present in at least one projection). Mathematically, for a reconruction $x$ and grount truth attenuation $\mu$,
\begin{align}
\label{eq.ploss}
L_p(x, \mu) &:= \| x - \mu \|_{\text{FullFoV}} + \alpha_1 (1.0 - \text{\texttt{SSIM}}_{\text{FullFoV}}(x, \mu)) + \nonumber \\
  & + \alpha_2 \| x- \mu \|_{\text{PartFoV}} + \alpha_2 \alpha_1 (1.0 - \text{\texttt{SSIM}}_{\text{PartFoV}}(x, \mu)),
\end{align}
where $\alpha_1 = 0.5$ and $\alpha_2$ was set to $0.1$ initially and then reduced to $0.01$ after the first learning rate decay step in order to prioritize reconstruction of the full field of view region. The projection domain loss is a weighted mean absolute error between scatter-corrected projection $y$ and the primary signal $y_{\text{primary}}$ from \eqref{eq.normprimary}, i.e.,
\begin{equation*}
L_d(y, y_{\text{primary}}) := 10 \| y - y_{\text{primary}}\|.
\end{equation*}
Reconstruction losses for all $x \in I$ are computed and summed. As a data augmentation strategy, we randomply flipped along the left-right and the head-foot axes. Isocenter was chosen by adding a random offset sampled from an isotropic Gaussian distribution with $0$ mm mean and a standard deviation of $100$ mm to the volume center. 

LIRE++ was trained to reconstruct complete volumes. NVIDIA H100 GPUs with gradient accumulation were used to achieve effective batch size of $8$. Adam optimizer \cite{kingma2014} was employed with an initial learning rate of $0.001$ and a plateau scheduler with linear warm-up and 10 epoch patience. At the end of each epoch models were evaluated, the best model was picked for testing. During training, LIRE++ used around 50 GB of GPU memory with internal patch size of 128x128x128, however, using smaller patch size can keep the GPU memory usage under 24 GB with identical reconstruction and parameter gradients. For comparison, $\partial$U-Net does not have this flexibility in GPU memory usage and always requires around 48 GB of memory during training.

In order to demonstrate how LIRE++ can be scaled to 1 mm data, we have added another primal/dual block on top of a pre-trained 2 mm version of LIRE++ and finetuned the whole network on simulated pelvic data. The added primal block uses reduced base filter count of $24$ and the dual block base filter count is $64$. This extended version of LIRE++ for 1 mm reconstruction can be trained on GPUs with 48 GB memory using smaller patch sizes and CPU-GPU streaming for latent vectors. During inference, the GPU memory utilization of the exteded LIRE++ remains under 24 GB.

\section{Results}
\label{s.results}

\subsection{Image quality: synthetic data}
\label{s.synthdata}

\begin{table*}[t]
\caption{Test results on simulated CBCT data, best result in bold; mean $\pm$ std.dev. for each metric. Mean inference time in seconds, parameter count in millions.}
\label{tab:met-all}
\centering
\begin{tabular}{|c|l l l| c c |}
\hline
  Method & \multicolumn{1}{c}{PSNR} & \multicolumn{1}{c}{SSIM} & \multicolumn{1}{c|}{MAE} & time  & par.  \\
  & & & \multicolumn{1}{c|}{(HU)} & (sec.) & (M) \\
  \hline
  \multicolumn{6}{|c|}{Thorax} \\ 
  \hline
FDK & $18.42 \pm 2.10$ & $0.65 \pm 0.06$ & $251.06 \pm 33.62$ & $1$ & $7.8$\\
TV & $31.76 \pm 2.08$ & $0.89 \pm 0.03$ & $53.02 \pm 6.51$ & $600$ & $7.8$\\
U-Net & $37.75 \pm 2.08$ & $0.92 \pm 0.02$ & $26.11 \pm 3.55$ & $3$ & $31.1$\\
$\partial$U-Net & $38.51 \pm 3.52$ & $0.96 \pm 0.01$ & $23.86 \pm 3.16$ & $3.5$ & $34.4$\\
LIRE   & $38.39 \pm 2.11$ & $0.96 \pm 0.01$ & $24.17 \pm 3.33$ & $30.5$ & $32.2$\\
LIRE+   & $38.32 \pm 2.10$ & $0.96 \pm 0.01$ & $24.49 \pm 3.31$ & $14.7$ & $17.1$\\
LIRE++  & $\mathbf{39.56 \pm 2.18}$ & $\mathbf{0.97 \pm 0.01}$ & $\mathbf{21.68 \pm 3.13}$ & $7.3$ & $15.8$\\
  \hline
  \multicolumn{6}{|c|}{Pelvic \& Abdominal} \\
    \hline
FDK & $17.72 \pm 5.40$ & $0.74 \pm 0.04$ & $305.08 \pm 29.33$ & $1$ & $7.8$\\
TV & $33.23 \pm 5.30$ & $0.73 \pm 0.17$ & $53.92 \pm 4.88$ & $600$ & $7.8$\\
U-Net & $40.16 \pm 5.67$ & $0.87 \pm 0.08$ & $21.35 \pm 3.95$ & $3$ & $31.1$\\
$\partial$U-Net & $39.56 \pm 5.92$ & $0.87 \pm 0.09$ & $23.52 \pm 5.35$ & $3.5$ & $34.4$\\
LIRE++  & $\mathbf{41.73 \pm 6.53}$ & $\mathbf{0.90 \pm 0.07}$ & $\mathbf{19.74 \pm 5.68}$ & $7.3$ & $15.8$\\
\hline
\end{tabular}
\end{table*}

\begin{figure}[h!]
    \centering
    \includegraphics[width=0.95\linewidth]{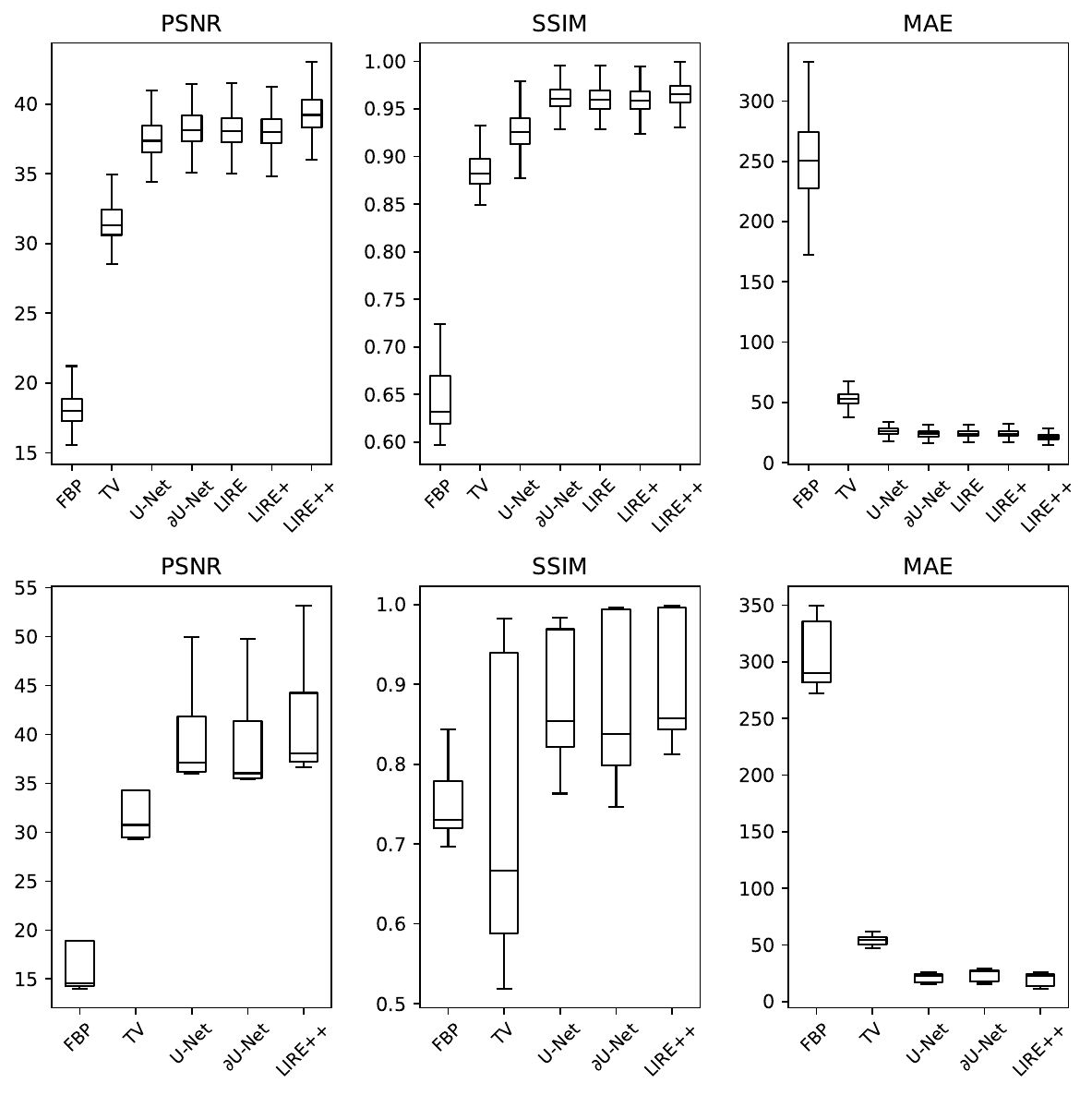}
    \caption{Reconstruction quality metrics. Thorax in the top row, pelvic \& abdominal in the bottom row. }
    \label{fig:met-all}
  \end{figure}

  \begin{figure}[h!]
    \centering
    \subfloat[]{\includegraphics[width=0.25\linewidth]{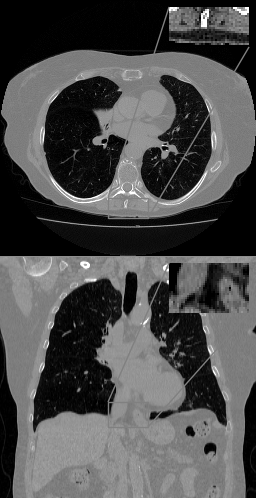}}
    \hfil
    \subfloat[]{\includegraphics[ width=0.25\linewidth]{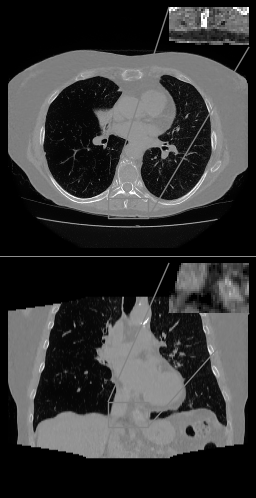}}
    \hfil
    \subfloat[]{\includegraphics[ width=0.25\linewidth]{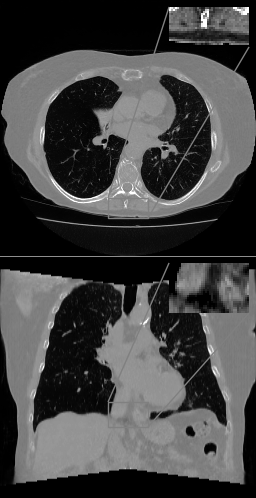}}
    \hfil
    \subfloat[]{\includegraphics[ width=0.25\linewidth]{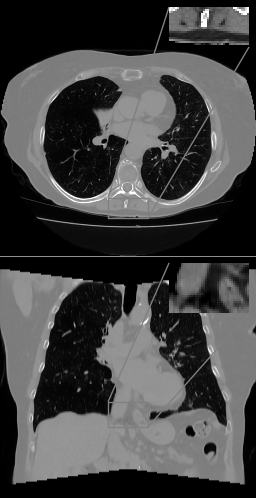}}
    \caption{(a) Axial (top) and coronal (bottom) slices of thorax CT, HU range=(-1000, 800) and (-150, 250) for ROI, (b) U-net (c) $\partial$U-net, (d) LIRE++}
    \label{fig:all-thorax}
\end{figure}

\begin{figure}[h!]
    \centering
    \subfloat[]{\includegraphics[width=0.25\linewidth]{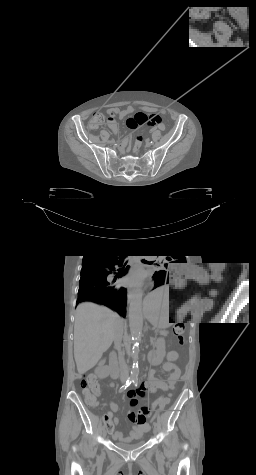}}
    \hfil
    \subfloat[]{\includegraphics[width=0.25\linewidth]{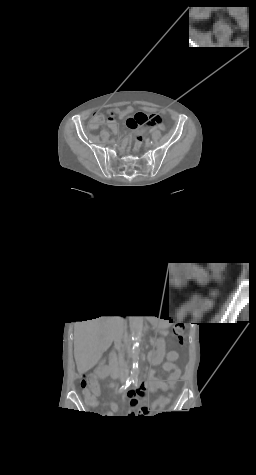}}
    \hfil
    \subfloat[]{\includegraphics[width=0.25\linewidth]{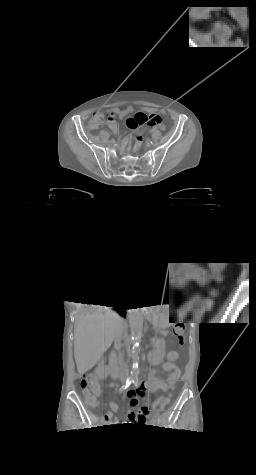}}
    \hfil
    \subfloat[]{\includegraphics[width=0.25\linewidth]{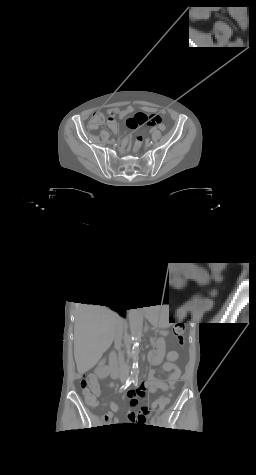}}
    \caption{(a) Axial (top) and coronal (bottom) slices of abdominal CT, HU range=(-400, 400) and (-150, 250) for ROI, (b) U-net (c) $\partial$U-net, (d) LIRE++}
    \label{fig:all-pelvic}
\end{figure}

We perform extensive evaluation of LIRE++ and the baselines using image quality metrics such as PSNR and SSIM, which are computed for attenuation values, as well as MAE in Hounsfield Units due its importance for radiotherapy applications.

In Table \ref{tab:met-all} we report these metrics on the thorax \& pelvic test set, and the corresponding box plots are provided in Figure \ref{fig:met-all}. All metrics are computed for the full field of view region, i.e., the voxels which are present in at least half of the projections, which coincides with the field of view given by FDK and TV methods. Table \ref{tab:met-all} also contains mean total inference times per volume on NVIDIA A100 accelerator and the parameter counts, where in case of FDK and TV the parameter count of scatter pre-correction U-net is provided. In case of TV reconstruction, high inference time is partially due to multiple CPU-GPU memory transfers in ODL. Examples of thorax image slices of a ground truth image and the corresponding reconstructions from baselines and LIRE++ are presented in Fig. \ref{fig:all-thorax}.  Similarly, pelvic \& abdominal image slices are presented in Fig. \ref{fig:all-pelvic}. The image samples demonstrate particularly well that LIRE++ is superior in reproduction of these soft tissue details which appear blurred in the baselines. Field-of-view in the reconstructions given by LIRE++ and $\partial$U-net is increased since the training loss is optimized over all voxels which are present in at least one projection. Extended FoV reconstruction quality for the voxels which are observed in at least one projection, but less than half of all projections, is slightly higher in LIRE++ reconstructions compared to $\partial$U-Net by appoximately 1 dB higher PSNR.

Compared to LIRE and LIRE+, LIRE++ is superior as well.  However, both LIRE and LIRE+ were finetuned on scatter pre-corrected data instead of being trained from scratch, which can have a negative impact on the reconstruction quality. Additionally, even though LIRE/LIRE+ support gradient computation for the projection data, we disabled it for consistency with other baselines and the lack of end-to-end trained scatter correction in LIRE/LIRE+ could be detrimental as well.

\subsection{Image quality: real data}
\label{s.realdata}

\begin{table*}[t]
\caption{Mean ROI intensity difference on real data}
\label{tab:met-real}
\centering
\begin{tabular}{|c|c c c c| c |}
  \hline
  & \multicolumn{4}{|c|}{Mean ROI difference (HU)} & MAE\\
Method & Fat & Muscle & Bone  & Bladder & (HU) \\
  \hline
FDK & $-341$ & $-323$ & $-497$ & $-185$ & $118$\\
FDK (cal.) & $-59$ & $-99$ & $-142$ & $84$ & $91$\\
TV++ & $-1$ & $-42$ & $-59$ & $7$ & $65$\\
LIRE++ & $-37$ & $-41$ & $-30$ & $12$ & $56$\\
\hline
\end{tabular}
\end{table*}

\begin{figure}[t]
    \centering
    \subfloat[]{\includegraphics[width=0.25\linewidth]{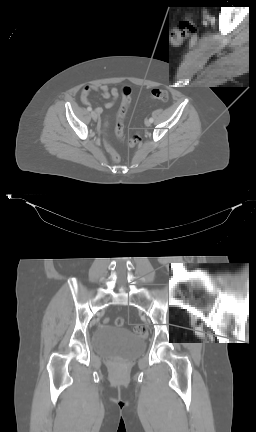}}
    \hfil
    \subfloat[]{\includegraphics[ width=0.25\linewidth]{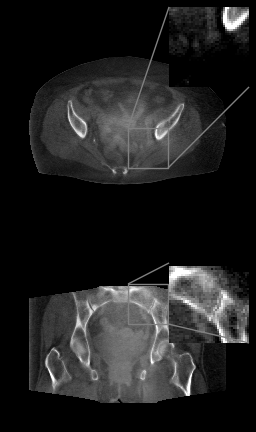}}
    \hfil
    \subfloat[]{\includegraphics[ width=0.25\linewidth]{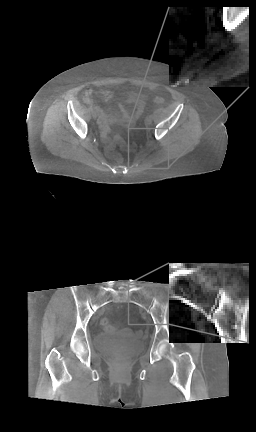}}
    \hfil
    \subfloat[]{\includegraphics[ width=0.25\linewidth]{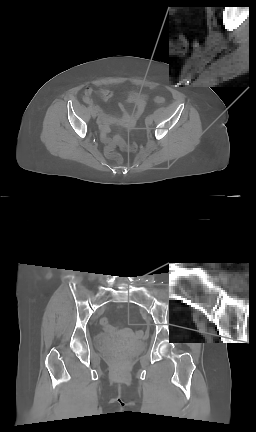}}
    \caption{(a) Axial (top) and coronal (bottom) slices of planning pelvice CT, HU range=(-400, 400) and (-150, 250) for ROI, (b) FDK (c) TV++ (d) LIRE++}
    \label{fig:real-pelvic}
\end{figure}

\begin{figure}[t!]
    \centering
    \subfloat[]{\includegraphics[ width=0.5\linewidth]{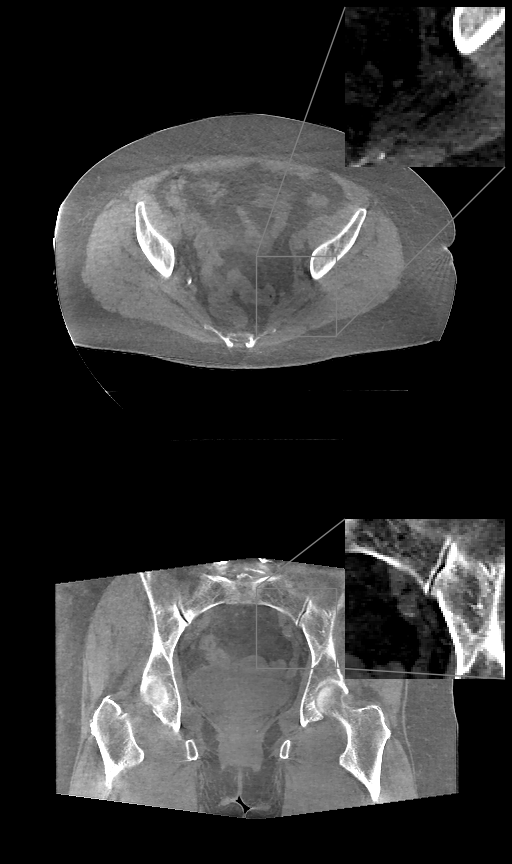}}
    \hfil
    \subfloat[]{\includegraphics[ width=0.5\linewidth]{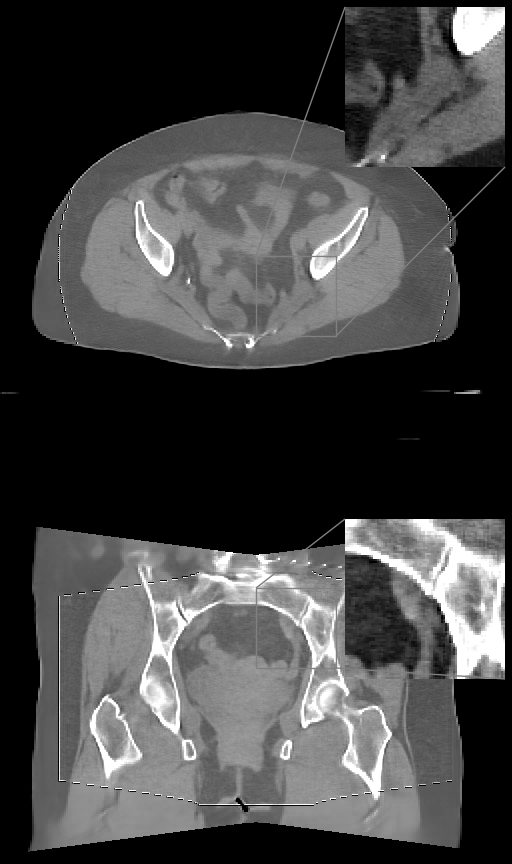}}
    \caption{(a) Axial (top) and coronal (bottom) slices of pelvic TV++ reconstruction, HU range=(-400, 400) and (-150, 250) for ROI, (b) LIRE++}
    \label{fig:real-pelvic-1mm}
  \end{figure}
  
We perform a proof-of-concept evaluation of LIRE++ on real CBCT pelvic data and compare it to the FDK baseline with U-net for scatter pre-correction and a proprietary TV++ method currently in use in our center. In addition to FDK we evaluate its calibrated version, where the HU values from FDK reconstruction all undergo a single affine transformation, which was determined by a linear regression matching central slices of the FDK reconstructions with the corresponding slices of planning CTs. To provide quantitative a comparison in terms of HU accuracy, we used planning CT and rigid registration.

The results are presented in Table \ref{tab:met-real}. MAE in Hounsfield Units is measured in the central full field of view region. Additionally, we selected four spherical regions of interest between 2 and 4 cm in diameter, which are well aligned in planning and CBCT, and computed the mean HU intensities inside these regions to measure reproduction accuracy of various regions. The mean difference of these HU averages between planning CT and the reconstructions are given in Table \ref{tab:met-real} as well. Axial and coronal image slices are presented in Figure \ref{fig:real-pelvic}.

This comparison demonstrates that LIRE++ translates well to real CBCT pelvic data. Reconstruction given by LIRE++ is noticeably cleaner than the TV++ reconstruction, scatter artifacts in particular are well-suppressed. Field of view given by LIRE++ is slightly larger compared to TV++. We have measured an improvement in mean HU accuracy, however, due to anatomical differences such comparison can underestimate actual reconstruction quality. TV++, on the other hand, substantially outperforms a classical FDK method with deep-learning scatter precorrection and its calibrated version.

In order to demonstrate that LIRE++ can be scaled to full resolution, we provide sample reconstructions with 1 mm voxel pitch in Figure \ref{fig:real-pelvic-1mm} from the extended version of LIRE++ and compare them to TV++ reconstructions. The reconstructions from LIRE++ are less noisy; however, more finetuning might be needed to completely remove image artifacts.

\section{Discussion}
\label{s.discuss}

We have introduced LIRE++, trained it on synthetic CBCT data and evaluated using synthetic as well as real CBCT data. On synthetic data, we have observed noticeable improvements over the baselines for both thorax and abdominal/pelvic datasets. Scatter-induced artifacts are well-suppressed in spite of their non-local nature and the absence of self-attention layers in LIRE++. The new model combines the multi-scale approach of LIRE+ and the U-net architecture in primal blocks from LIRE, resulting in a large receptive field in the primal domain, which can be particularly helpful for scatter correction. Furthermore, LIRE++ translates well to real pelvic CBCT acquisitions, where it compares favourably with a proprietary state-of-the-art reconstruction method. Given reasonable inference time of around 7 seconds at 2 mm voxel pitch and around 40 seconds at 1 mm voxel pitch on NVIDIA A100 GPU, LIRE++ has the potential to replace classical reconstruction methods in pelvic CBCT radiotherapy applications, where the extended field of view provided by LIRE++ could be of interest in particular.

Nevertherless, there remain potential extensions of our study for future research. Firstly, we evaluate LIRE++ on real projections using pelvic data only, since thorax CBCT scans in our center are always acquired with anti-scatter grids installed which we do not simulate at the moment. Additionally, the projection count for phase-resolved thorax CBCT is lower, and the field of view is typically set to the `small' setting. Therefore, a dedicated version of LIRE++ would be desirable for phase-resolved thorax CBCT scans, however, architectural changes are not strictly needed.

Secondly, LIRE++ performs well as a 3D reconstruction method, but we do not handle motion-induced artifacts at the moment. Directly incorporating some form of motion compensation in LIRE++ in order to obtain a complete 4D reconstruction is an interesting research direction.

\newpage
\section*{Acknowledgements}
This work is funded by grants from Elekta AB and the Netherlands Enterprise Agency (PPS2102). The Radiotherapy Department of Netherlands Cancer Institute (NKI) receives royalties for Cone Beam CT radiotherapy software from Elekta AB.

The authors would also like to acknowledge the support of Research High Performance Computing (RHPC) facility of the Netherlands Cancer Institute.

\bibliographystyle{elsarticle-harv} 
\bibliography{bibliography}




\newpage

\section*{Appendix}
\label{ss.physics}
A photon travelling through human tissues at typical X-ray energies in the $20-120$ kEv range can either pass through unhindered, or undergo one of the following most common interactions:
\begin{enumerate}
\item photoelectric absorbtion, where the photon is absorbed and an electron is ejected;
\item Compton scattering, where the photon collides with an electron, causing the electron to recoil and a scattered photon with lower energy to be emitted;
\item Rayleigh scattering, where the photon interacts with the whole atom, and a scattered photon with the same energy is emitted.
\end{enumerate}
Occurence of any of these interactions, as well as the direction of scattered photon in case of Compton and Rayleigh interactions, is probabilistic in nature and depends on the atomic composition of the material and the photon energy. For the purposes of this paper we assume that patients are composed from water and bone materials, i.e., each voxel is a mix of bone and water densities. To specify this decomposition, we will use dimensionless relative densities $\rho^w(x)$ and $\rho^b(x)$ which measure the density of material $m \in \{ w, b\}$ present in a voxel $x$ relative to the density of material $m$ under normal conditions.

Firstly, consider a photon $\gamma$ with energy $e>0$ traveling from its initial position $x \in \mathbb R^3$ in the direction of unit vector $\vec v \in S^2$, $S^2:=\{\vec v \in \mathbb R^3: \| \vec v\| = 1\}$. Then, according to Beer-Lambert law, the probability $p_0(x, \vec v, e)$ that the photon escapes the patient is given by
\begin{equation}
\label{eq.beer}
p_0(x, \vec v, e) := \exp\left( - \int_0^{\infty} \mu_{\text{tot}, e}(x + t \vec v) \mathrm d t\right).
\end{equation}
Here, $\mu_{\text{tot}, e}(\cdot): \Omega_X \to \Rp$ is a function specifying \emph{total attenuation coefficient}, measured in $\mathrm{mm}^{-1}$ in this paper, at a given energy level $e > 0$, measured in keV, in the spatial domain $\Omega_X \subset \mathbb R^3$ occupied by the patient.  A photon which passes through the patient unhindered and which is recorded by the X-ray detector is called a \emph{primary} photon. More generally, Beer-Lambert law implies that
\begin{equation}
\mathbb P(\gamma \text{ travels distance} > s| x, \vec v, e) = \exp\left( - \int_0^s \mu_{\text{tot}, e}(x + t \vec v) \mathrm d t\right), 
\end{equation}
therefore, for $s \geq 0$
\begin{align}
\label{eq.lambdameas}
  &\lambda (s | x, \vec v, e):=\mathbb P(\gamma \text{ interacts in } [0, s]| x, \vec v, e) = \nonumber \\
  &= 1 - \exp\left( - \int_0^s \mu_{\text{tot}, e}(x + t \vec v) \mathrm d t\right). 
\end{align}
Under reasonable assumptions about $\mu_{\text{tot}, e}$, $\lambda$ is continuous, non-decreasing and can be used to define integrals w.r.t. photon travel distance as Lebesgue-Stieltjes integrals $\int f(s) d \lambda(s|x, \vec v, e)$ \citep{halmos1974}. In order to conditionally sample interaction distance $u \sim \lambda$ via inverse transform method it suffices to sample a uniform random variable $\overline u \sim \mathcal U(0,1)$ and solve the equation
\begin{equation}
\label{eq.distsample}
\lambda (u | x, \vec v, e) = (1 - p_0(x, \vec v, e)) \overline u \quad \text{for } u>0,
\end{equation}
in this case we write $u \sim_{\overline u} \lambda$. For $u \sim \lambda$, a single-sample Monte Carlo estimate of an integral $\int f(s) d \lambda(s|x, \vec v, e)$ with respect to the travel distance is given by
\begin{equation}
\label{eq.mcdist}
\int f(s) d \lambda(s|x, \vec v, e) \approx (1 - p_0(x, \vec v, e)) f(u).
\end{equation}

Secondly, given water-bone decomposition, total attenuation can be decomposed as $\mu_{\text{tot}, e} = \mu_{\text{tot}, e}^w + \mu_{\text{tot}, e}^b$ for the corresponding water and bone attenuation components, where $ \mu_{\text{tot}, e}^w, \mu_{\text{tot}, e}^b: \Omega_X \to \Rp$. For $m \in \{ w, b\}$ and $x \in \mathbb R^3$, $\mu_{\text{tot}, e}^{m}(x) = \rho^{m}(x) \overline \mu_{\text{tot}, e}^{m}$ where $\overline \mu_{\text{tot}, e}^{m} \in \Rp$ is the total attenuation of $m$ at energy $e>0$ under normal temperature and pressure conditions. If the photon undergoes an interaction at a point $x$ along its path, the conditional probability that it has interacted with material $m \in \{w, b\}$ is given by 
\begin{equation}
  \mathbb P(\text{interaction with } m|\text{photon interacted at } x) = \frac{\mu_{\text{tot}, e}^m(x)}{\mu_{\text{tot}, e}(x)}.
\end{equation}
These considerations will allow to reduce sampling scattered photon paths for a mix of materials to a hierarchical sampling procedure wherein interacting material is sampled from a Bernoulli distribution first, so in the remainder of the section we focus on modeling a single material and omit it in the notation.

Thirdly, if a photon has interacted at a point $x$, the conditional probabilities for each specific interaction type can be determined, since the total attenuation coefficient $\mu_{\text{tot}, e}$ for a particular material is composed from the corresponding photoelectric (p), Compton (c) and Rayleigh (r) attenuation components:
\begin{equation}
\mu_{\text{tot}, e} = \mu_{\text{p}, e} + \mu_{\text{c}, e} + \mu_{\text{r}, e},
\end{equation}
where $\mu_{\text{p}, e}, \mu_{\text{c}, e}, \mu_{\text{r}, e}: \Omega_X \to \Rp$. Then the conditional probability that a specific interaction $\text{T} \in \{p, c, r\}$ took place can be computed as
\begin{equation}
  \mathbb P(\text{interaction type T}|\text{photon interacted at } x) = \frac{\mu_{\text{T}, e}(x)}{\mu_{\text{tot}, e}(x)}.
\end{equation}
The relative frequencies of photoelectric, Compton and Rayleigh events depend on the photon energy and the atomic composition of the material, and in practice this cross-section data is available for many standard materials such as water and cortical bone in specialized databases. We rely on the xraylib library \citep{xraylib2011} to access this information.

Finally, to specify scatter distribution, it is necessary to define a conditional measure $\nu(\cdot |x, \vec v, e)$ on $S^2 \times \Rpp$ which determines scattering direction and energy of photon which has interacted at a point $x$ with initial direction $\vec v$ and energy $e$. $\nu$ is in general not a probability measure, since $ \nu( S^2 \times \Rpp |x, \vec v, e)$ by definition equals the conditional probability that the photon which has interacted has undergone either Compton or Rayleigh scattering. It is known that the distribution defined by $\nu$ on $S^2$ is invariant w.r.t. rotations along $\vec v$. Additionally, if a photon with energy $e > 0$ undergoes Compton scattering with scatter angle $\theta$ between new and old directions, the energy $e'$ of the scattered photon is reduced and is given by
\begin{equation}
e' = \frac{e}{1 + \frac{e}{m_e c^2} (1 - \cos \theta)}.
\end{equation}
If, on the other hand, a photon undergoes Rayleigh scattering, its energy remains unchanged and $e' = e$. Therefore, the measure $\nu$ can be completely determined from the \emph{differential cross-section} data for the scattering angle $\theta$ for Compton and Rayleigh interactions, which can be accessed via e.g. xraylib library, and the formulas for energy above.

\begin{algorithm}
\centering
\begin{algorithmic}[1]
  \Procedure{\texttt{SamplePath}}{$\mathit{Src}$, $\mu$}
\State $P \gets []$ \Comment{Initialize output list}
\State $S \gets \mathrm{Sobol}_{5n}(N)$ \Comment{Get $N=|\mathit{src}|$ samples of $5n$-dimensional Sobol sequence}
\For{$i \gets 1, \dots, N$}
\State $x, \vec v, e \gets x_0, \mathit{Src}[i][0], \mathit{Src}[i][1]$ \Comment{Get source direction \& energy} 
\State $X, E \gets [x], [e]$ \Comment{Init lists of positions \& energies}
\State $l \sim_{S[i][0]} \lambda(\cdot |x,\vec v,e)$ \Comment{Sample interaction distance}
\State $w \gets 1-p_0(x, \vec v, e)$ \Comment{$\mathbb P (\text{photon doesn't escape})$} 
\State $W \gets [w]$ \Comment{Append weight} 
\State $x \gets x+l \vec v$ \Comment{Compute interaction point}
\State $X \gets X + [x]$ \Comment{Append interaction point}
\State $V \gets [\vec v]$ \Comment{List of direction vectors}
\For{$j \gets 1, \dots, n$}
\State $k \gets 1+5(j-1)$ \Comment{Offset for Sobol sequence}
\State $\mathrm{m} \sim_{S[i][k]} \mathrm{Ber}(\frac{\mu_{\text{tot}^, e}^w(x)}{\mu_{\text{tot}, e}(x)})$ \Comment{Sample material $\mathrm{m} \in \{w, b\}$}
\State $\mathrm{T} \sim_{S[i][k+1]} \mathrm{Ber}(\frac{\mu_{\text{c}, e}^{\mathrm m}(x)}{\mu_{\text{c}, e}^{\mathrm m}(x) + \mu_{\text{r}, e}^{\mathrm m}(x)})$ \Comment{Sample interaction $\mathrm{T} \in \{c, r\}$}
\State $w \gets w \cdot \frac{1-\mu_{\text{p}, e}^{\mathrm m}(x)}{\mu_{\text{tot}, e}^{\mathrm m}(x)}$ \Comment{$\mathbb P ( \mathrm{T} \in \{c, r\} | \text{photon interacts})$}
\State $\vec v, e \sim_{S[i][k+2], S[i][k+3]} \nu_{\mathrm m, \mathrm T}(\cdot |\vec v, e)$ \Comment{Sample direction \& energy}
\State $l \sim_{S[i][k+4]} \lambda(\cdot |x,\vec v,e)$ \Comment{Sample interaction distance}
\State $w \gets w (1-p_0(x, \vec v, e))$ \Comment{$\mathbb P (\text{photon doesn't escape})$}
\State $W \gets [w]$ \Comment{Append weight} 
\State $x \gets x+l \vec v$ \Comment{Compute interaction point}
\State $X \gets X + [x]$ \Comment{Append interaction point}
\State $V \gets V + [\vec v]$ \Comment{Append direction vector}
\State $E \gets E + [e]$ \Comment{Append energy}
\EndFor
\State $P \gets P + [X, V, E, W]$
\EndFor
\State \textbf{return} $P$
\EndProcedure
\end{algorithmic}
\caption{Path sampling}
\label{alg:scatter}
\end{algorithm}

\begin{algorithm}
\centering
\begin{algorithmic}[1]
\Procedure{\texttt{Integrate}}{$P$, $\mu$}
\State $S \gets \mathrm{zeros}(\text{Detector})$ \Comment{Initialize zero scatter estimate}
\State $\vec n \gets \text{DetectorNormal}$ \Comment{Get detector normal}
\For{$[X, V, E, W] \in P$} \Comment{Loop over paths}
\For{$\sigma \in \text{Detector}$} \Comment{Loop over detector elements $\sigma$}
\For{$[x, \vec v, e, w] \in \mathrm{zip}(X, V, E, W)$} \Comment{Loop over inter. points}
\State $\vec v_s \gets \mathrm{normalize}(\sigma_{\text{center}} - x)$ \Comment{Vector to pixel center}
\State $e' \gets \frac{e}{1 + e(1 - \langle \vec v, \vec v_s \rangle)/(m_e c^2)}$ \Comment{Compton scattering energy}
\State $s_c \gets \nu_{w, c} \frac{\mu_{\text{tot}^, e}^w(x)}{\mu_{\text{tot}, e}(x)} + \nu_{b, c} \frac{\mu_{\text{tot}^, e}^b(x)}{\mu_{\text{tot}, e}(x)}$
\State $s_r \gets \nu_{w, r} \frac{\mu_{\text{tot}^, e}^w(x)}{\mu_{\text{tot}, e}(x)} + \nu_{b, r} \frac{\mu_{\text{tot}^, e}^b(x)}{\mu_{\text{tot}, e}(x)}$
\State $s \gets s_c p_0(x, \vec v_s, e') \mathrm{resp}(e') + s_r p_0(x, \vec v_s, e) \mathrm{resp}(e)$
\State $S[\delta] \pluseq \frac{s  w |\langle \vec n, \vec v_s \rangle| \mathrm{area}(\sigma) }{\| \sigma_{\text{center}} - x\|_2^2}$ \Comment{Approximate $\int\limits_{\mathrm{proj}_{x}(\sigma)}$}
\EndFor
\EndFor
\EndFor
\State \textbf{return} $S$
\EndProcedure
\end{algorithmic}
\caption{Path integration}
\label{alg:scatter2}
\end{algorithm}

\end{document}